\documentclass[aps,prd,preprint,superscriptaddress,showpacs,nofootinbib]{revtex4-1}
\pdfoutput=1
\usepackage{latexsym}
\usepackage{amssymb, amsmath}
\usepackage{epsfig, graphics,pstricks, color}
\numberwithin{equation}{section}
\usepackage{color}

\usepackage{booktabs} % for much better looking tables

\newcommand{\beq}{\begin{equation} }
\newcommand{\eeq}{\end{equation} }
\newcommand{\beqa}{\begin{eqnarray} }
\newcommand{\eeqa}{\end{eqnarray} }
\begin{document}

\maketitle
\flushbottom
\baselineskip=21pt \pagestyle{plain} \setcounter{page}{1}

\vspace*{-1.1cm}

%\begin{flushright}{\small pub number}\end{flushright}

\vspace*{0.2cm}

\begin{center}

{\Large \bf   Neutrino Lighthouse at Sagittarius A*}\\ [9mm]

{\normalsize \bf Y. Bai$^1$, A. J. Barger$^{2,3,4}$, V. Barger$^1$, R. Lu$^1$, A. D. Peterson$^1$, J. Salvado$^{1,5}$}  \\ [4mm]
{\it
1. Dept. of Physics, University of Wisconsin, Madison, WI 53706 \\
2. Dept. of Astronomy, University of Wisconsin, Madison, WI 53706 \\
3. Dept. of Physics and Astronomy, University of Hawaii, Honolulu, HI 96822 \\
4. Institute for Hawaii, University of Hawaii, Honolulu, HI 96822 \\
5. Wisconsin IceCube Particle Astrophysics Center, Madison, WI 53706}\\

\vspace*{0.5cm}

\today

\vspace*{0.7cm}

{\bf \small Abstract}

\vspace*{-0.2cm}
\end{center}
We investigate whether a subset of high-energy events observed by IceCube may be due to neutrinos from Sagittarius A*.  We check both spatial and temporal coincidences of IceCube events with other transient activities of Sagittarius A*. Among the seven IceCube shower events nearest to the galactic center, we have found that event 25 has a time very close to (around three hours after) the brightest X-ray flare of Sagittarius A* observed by the \emph{Chandra X-ray Observatory} with a $p$-value of 0.9\%. 
%Another weaker hint is that the brightest X-ray flare observed by the \emph{Swift} satellite occurred eight days after event 2. 
Furthermore, two of the seven events occurred within one day of each other (there is a 1.6\% probability that this would occur for a random distribution in time). Thus, the determination that some IceCube events occur at similar times as X-ray flares and others occur in a burst could be the smoking gun that Sagittarius A* is a point source of very high energy neutrinos. We point out that if IceCube galactic center neutrino events originate from charged pion decays, then TeV gamma rays should come from neutral pion decays at a similar rate. We show that the CTA, HAWC, H.E.S.S. and VERITAS experiments should be sensitive enough to test this.
\thispagestyle{empty}
\newpage
\setcounter{page}{1}

%-------------------------------------------------------------------------------------------------------------------------------
\section{Introduction}
%-------------------------------------------------------------------------------------------------------------------------------

The origins of Ultra High Energy (UHE) and Very High Energy (VHE) Cosmic Rays (CRs) and UHE and VHE astrophysical neutrinos (NUs)  are major unknowns; see, e.g.~\cite{Halzen:2014fpa, Olinto:2013efa, Blasi:2014roa, Sigl:2013gya}. These two fundamental open questions may be interconnected, since  accelerated protons from a source interact with protons in a surrounding gas and thereby produce charged pions that decay to NUs and neutral pions that decay to photons.   The questions may be answerable with the advent of detectors of at least area km$^2$ and volume km$^3$.  The Pierre Auger  \cite{Auger:2012yc} and Telescope Array (TA)  \cite{Abu-Zayyad:2013vza} experiments are probing the highest energy CRs ($>10^{18}$ eV), but so far a definitive association of UHE CRs with astrophysical sources has proved elusive.  The proton or iron primaries of CRs get deflected by magnetic fields, and this compromises the identifications of the source locations.  The TA data show a possible hot spot region on the sky, but at low significance \cite{Fang:2014uja}.  Leading astrophysical candidates for CRs are Active Galactic Nuclei (AGNs), Starburst Galaxies (SBGs) and Galaxy Mergers; see, e.g.~\cite{Kashiyama:2014rza,Anchordoqui:2014yva,Laha:2013lka, Winter:2013cla, Dar:2014, Gonzalez-Garcia:2013iha,Taylor:2014hya} for some recent discussion.   The standard model of CRs is shock acceleration by the Fermi mechanism.  The energetics of the shock is derived from the central engine, which in galaxies is the supermassive black hole (SMBH) that resides at the compact central region, so either AGNs or SBGs are highly plausible sources for the production of protons and NUs of extreme energies.

Although searches for UHE NUs have so far not resulted in any detections; see e.g.~\cite{Gorham:2010kv}, the IceCube (IC) experiment \cite{Aartsen:2013jdh,Aartsen:2014gkd,Karle} has observed a game-changing 36 VHE NU events with energies in the 30 TeV to 2 PeV range.  These IC events could well be the key to solving the origins of CRs, though it should be noted that even the highest NU energies of the IC events, $1-2$~PeV, are well below those predicted for cosmic NUs associated with UHE CRs \cite{Anchordoqui:2013dnh}.
The IC events provide strong evidence ($5.7\sigma$) for a non-terrestrial component of the neutrino flux, since a number of them have directions well beyond the galactic disc.  The IC data allow a brand new approach for understanding the physics of VHE NUs, CRs and possibly even dark matter  \cite{Anchordoqui:2013dnh}.

A blazar sample has been recently considered in the context of the IC events  \cite{Krauss:2014tna}. Assuming that the X-ray to gamma-ray emission originates in the photo-production of pions by accelerated protons, it was concluded that the integrated predicted NU luminosity of these particular sources is large enough to explain the two detected PeV events. Another suggestion is that BL Lacs and pulsar wind nebulae may be the astrophysical counterparts of IC events \cite{Padovani:2014bha}.

Gamma-Ray Bursts (GRBs) are the most energetic electromagnetic events in the Universe.  They are extragalactic, and the bursts are of short duration, from 10 milliseconds to a few minutes.   Thus, it is natural to see if there are associated occurrences of NU bursts.  Tests were made by IC to see if VHE NU events, both showers and muon tracks, were  associated with known GRBs, but no evidence was found for any GRB coincidences \cite{Aartsen:2013jdh, Aartsen:2014gkd, Casey:2013, Yacobi:2014vja}. The analysis looked for temporal and spatial correlations with the GRBs reported by the \emph{Fermi} Gamma-ray Burst Monitor and \emph{Swift}.

Other astrophysical sources that are candidates for CRs and VHE NUs include TeV gamma-ray sources, HyperNovae (HNe), and SuperNovae (SNe).  The occurrence of HNe and SNe is high in SBGs, which makes them prime possibilities.  However, no significant association of IC events with SBGs has been established  \cite{Aartsen:2014gkd}.  Magnetars, pulsars, stellar black holes and binary systems are other astrophysical sources that have been considered in the search for a connection with the IC events.

Decays of Super Heavy Dark Matter (SHDM) with a very long lifetime could alternatively be the source of the IC events \cite{Kolb:1998ki,Chung:1998zb,Feldstein:2013kka,Bai:2013nga,Esmaili:2013gha,Zavala:2014dla, Esmaili:2012us}.  It  may not be easy to differentiate NUs from SHDM decays from those of astrophysical origin, because both should be proportional to the SHDM density to a first approximation.

In this paper, we advance arguments that Sgr A* is the source of a subset of IC NU events. In Section~\ref{ICevents}, we summarize salient characteristics of the IC new physics signal. In Section~\ref{GCNU}, we remark on the positional coincidence of 7 events with the galactic center.  In Section~\ref{flares}, we show that there is a suggestive time correlation of IC events with large flares at Sgr A*.  We find such correlations in {\em Swift\/}, {\em Chandra\/}, and {\em NuSTAR\/} X-ray data but not with \emph{Fermi} low-energy gamma rays. In Section~\ref{clustering}, we examine the time sequence of IC events in the GC and compute $p$-values of random explanations of the timing of the IC events using self-clustering analysis and a friends-of-friends clustering analysis, and we do a likelihood analysis for {\em Chandra\/} flare coincidence with IC events. A prediction of a proton-proton origin of the GC IC events is that there will be similar numbers of high-energy gamma ray events as IC NU events. In Section~\ref{gammaray}, we present our prediction, which can be confirmed with upcoming gamma-ray detection experiments \cite{Holder:2014eja}. In Section~\ref{conclusions} we give our conclusions.

%-------------------------------------------------------------------------------------------------------------------------------
\section{IceCube High-Energy Events}
\label{ICevents}
%-------------------------------------------------------------------------------------------------------------------------------

The IC experiment has amassed a very large dataset of atmospheric (ATM) NU events.  At energies $>1$~TeV, the ATM NU energy spectrum, $dN/dE_\nu$, falls approximately as $E_\nu^{-2.7}$.  Above 30~TeV, the observed number of NU events significantly exceeds an extrapolation of the ATM NU flux.   Above 300~TeV, the NU flux has an energy dependence consistent with $E_\nu^{-2}$ with a cutoff at a few PeV, or an energy dependence with $E_\nu^{-2.3}$ without a cutoff \cite{Aartsen:2014gkd,Anchordoqui:2013dnh}.  It has been suggested that the violation of Lorentz invariance could be a cause for the termination of the NU energy spectrum at a very high energy \cite{Anchordoqui:2014hua}.

The first IC search for high-energy NU events above the steeply falling ATM background has been based on the selection of events that start inside the detector (so called ``contained events").  The ATM background is rejected by veto of events in which a muon enters the detector at the same time  \cite{Gaisser:2014bja}, thus giving $4\pi$ coverage of the sky. This approach preferentially selects electron-NU or tau-NU initiated events, although a few muon-NU events are also observed with an angular resolution less than a degree.  In these so-called high-energy starting events (HESE), the electromagnetic showers from primary electron-NUs or tau-NUs have degraded pointing accuracy, of order 15$^\circ$  and larger, but these events provide accurate deposited energy determinations.

For through-going muon-NU events, where the NU interaction occurs in rock outside the detector, the track of the produced muon will point back to the source with an angular resolution less than a degree.  The through-going data are still being analyzed.  The through-going muon-NU events will provide superior directional information, but they are more dependent on upward acceptance, which is highest for directions near the horizontal. The energy deposited in the detector may only be $1/4$  to $1/10$ of the incident NU energy  \cite{Karle}.  A connection of muon-NU events with bright astrophysical sources has not been found  \cite{Aartsen:2014gkd}.

\begin{figure}[htb]
\includegraphics[width=\textwidth]{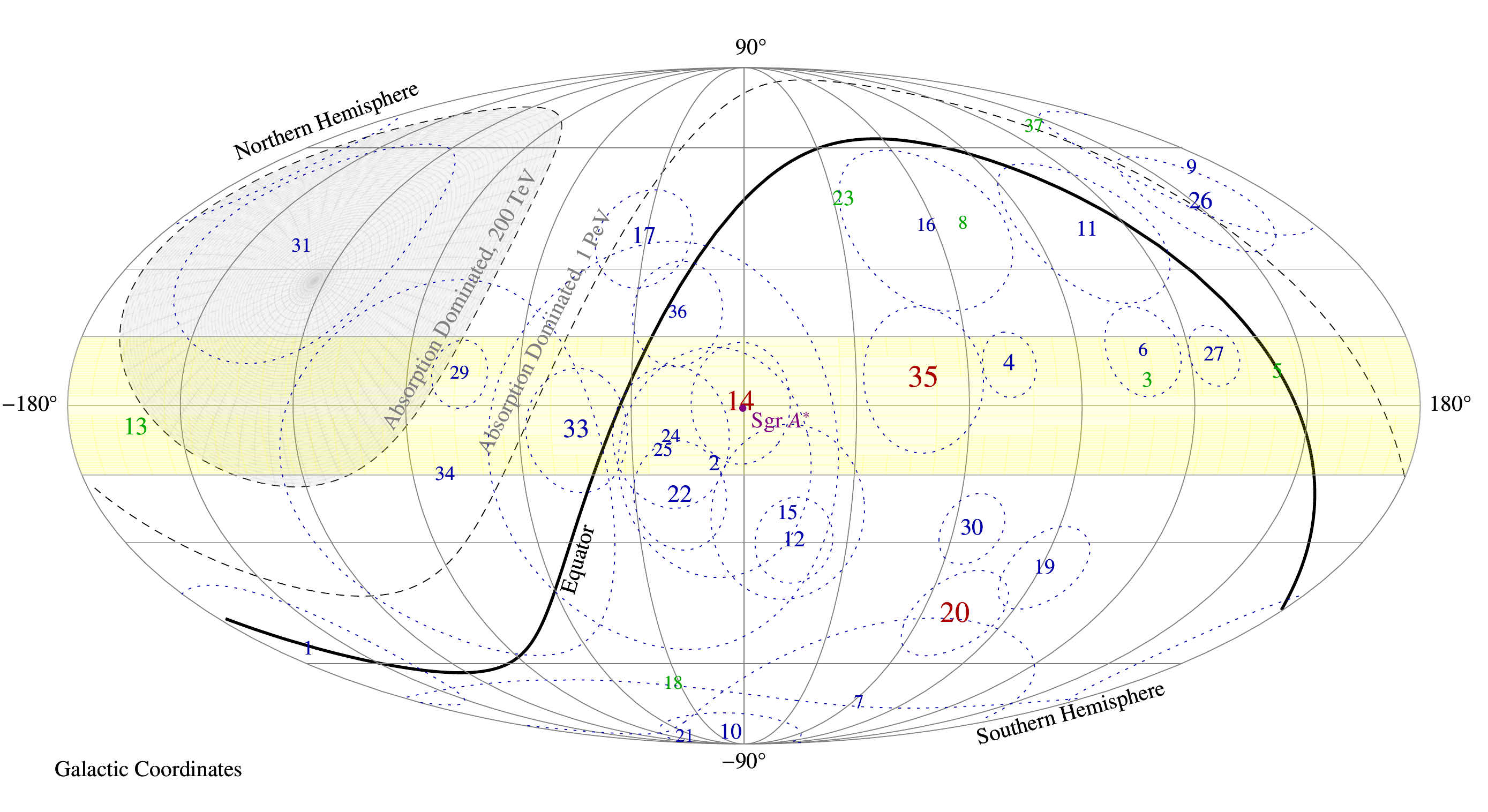}\\
\caption{\label{skymap} Skymap showing the IC NU events (labeled by event number; blue=shower, green=track, red=PeV shower) in galactic coordinates. The size of each NU event number label reflects the event's energy. High-energy NUs are absorbed by the Earth; the gray shading and dashed contours denote the regions where the flux attenuation is significant for $E_\nu = 200$~TeV and $E_\nu =1$~PeV \cite{Karle}. The black curve separates the Northern and Southern sky. The yellow shading denotes the galactic plane region. The dotted circles are $1\sigma$ angular regions for the IC events.
}
\end{figure}

The IC two-year data set has 28 events with in-detector deposited energies between 30~TeV and 1.1~PeV \cite{Aartsen:2013jdh}.  The three-year (988 days) data set consists of 36 events \cite{Aartsen:2014gkd}, well above the estimated backgrounds of $6.6\ (+5.9,-1.6)$ ATM NUs and $8.4 \pm 4.2$ ATM muons.  The significance of a new physics signal in the three-year data set in comparison to the ATM NU background is $5.7\sigma$, exceeding the nominal discovery criterion.  Most events are downward-going, because upward-going NUs suffer absorption by the Earth.  The three highest energy events are showers with energies of 1~PeV, 1.1~PeV and 2~PeV, all downward-going.  Thirty of the events are contained showers, and 6 events have a muon track.  The NU skymaps of these IC events show no significant clustering and are compatible with an isotropic distribution.  Moreover, the data are consistent with 1:1:1 NU flavor ratios, as would be expected from NUs coming from pion decays and their subsequent propagation as mass-eigenstates  \cite{Learned:1994wg,Beacom:2003nh}. We show a skymap of the three-year data in the galactic coordinate system (longitude, latitude) in Figure \ref{skymap}.

%-------------------------------------------------------------------------------------------------------------------------------
% FIGURE 1
%-------------------------------------------------------------------------------------------------------------------------------
%-------------------------------------------------------------------------------------------------------------------------------

%-------------------------------------------------------------------------------------------------------------------------------
% FIGURE 2
%-------------------------------------------------------------------------------------------------------------------------------
\begin{figure}[htb]
\includegraphics[width=\textwidth]{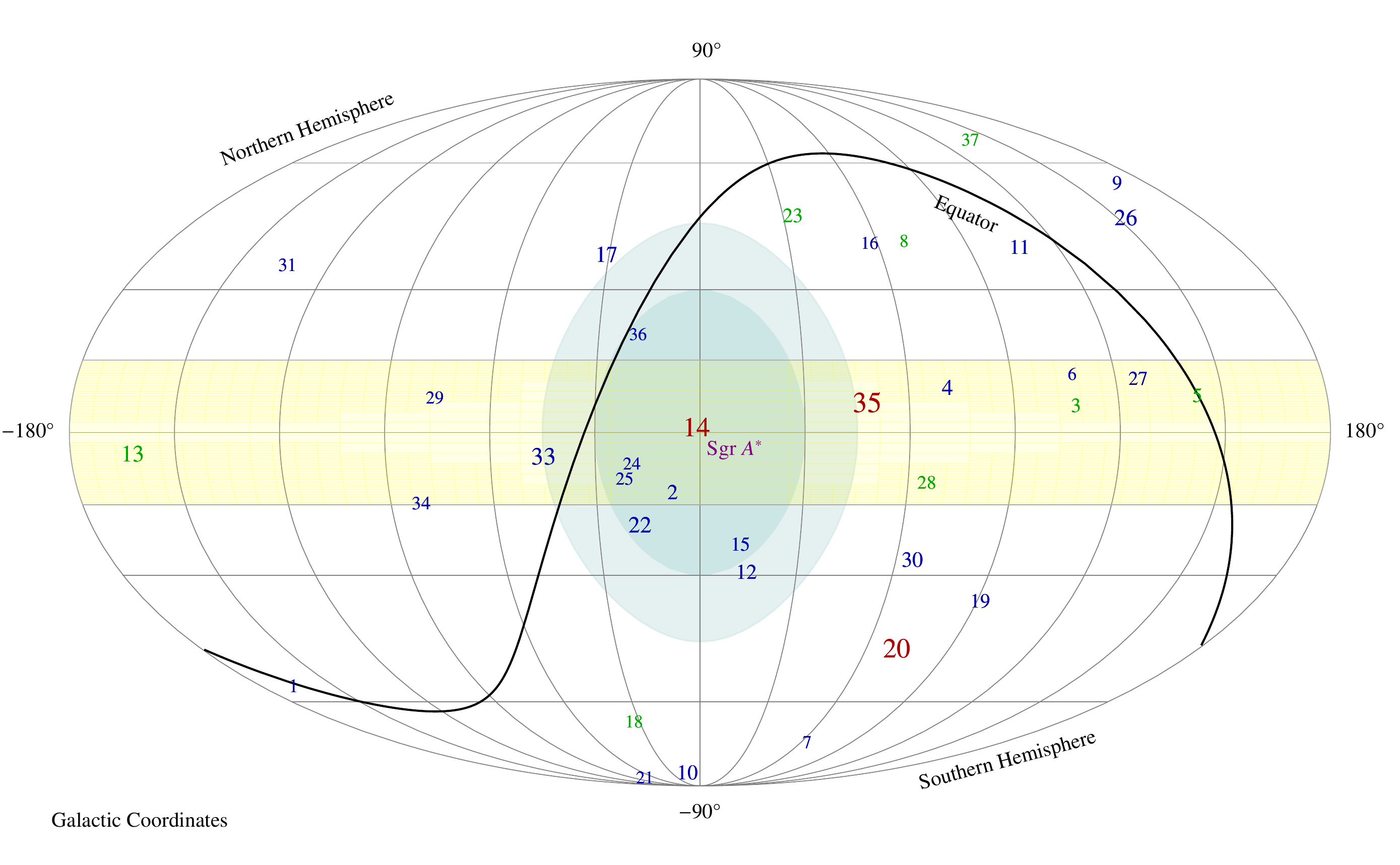}\\
\caption{\label{skymap2}
Skymap showing the IC NU events (labeled by event number; blue=shower, green=track, red=PeV shower) in galactic coordinates. 
The black curve separates the Northern and Southern sky. The yellow shading denotes the galactic plane region. The IC events that are positionally consistent with the GC are approximated by the dark (light) blue shaded region within $30^\circ$ ($45^\circ$) angular distance from the GC.
}
\end{figure}
%-------------------------------------------------------------------------------------------------------------------------------

The fact that the energy spectrum of the new physics NU signal shows a $E^{-2}$ dependence, from 60~TeV to 2~PeV, suggests that the Fermi shock mechanism is operative  \cite{Waxman:1998yy,Bahcall:1999yr}.  The Fermi mechanism would be applicable for either an SBG or an AGN  \cite{Stecker:2013fxa}.  The level of the IC NU flux is about $E_\nu^2\,dN_\nu/dE_\nu \approx 1.0\times10^{-8} \text{ GeV cm}^{-2}\text{ s}^{-1} \text{ sr}^{-1}$ per flavor \cite{Aartsen:2014gkd}.

%-------------------------------------------------------------------------------------------------------------------------------
\section{Galactic Center Neutrinos}
\label{GCNU}
%-------------------------------------------------------------------------------------------------------------------------------

The identification of astrophysical NU sources is a primary goal of NU telescopes, so evidence for an association of NU events with specific sources would be a huge scientific breakthrough.  VHE NUs may be galactic or extragalactic in origin and may well consist of a mixed composition of NU source types with differing NU energy spectra.  The high galactic latitudes of some of highest-energy IC events suggest at least some extragalactic component \cite{Aartsen:2014gkd}.

Point-source searches have been made by IC, both for CR events and for VHE NU events, to see if there are associations with galaxy clusters and other astrophysical objects (see Introduction).   The existence of many possible sources in a small region of the sky makes spatial-only identification difficult.  Indeed, IC events show no apparent evidence of spatial clustering of NU events, but they also do not exclude that possibility, because of the imperfect resolution on the pointing of the more numerous shower events.

Transient outbursts may be the most likely to yield the requisite extreme NU energies reported by IC.  However, IC searches for coincidences using three years of data, between April 2008 and May 2011, found no evidence for NU event coincidences with \emph{Fermi} gamma-ray data or with a selected catalog of binary systems and micro-quasars with known periodicities in X-ray, gamma-ray and radio data \cite{Aguilar:2013}. However, since the search for a time coincidence of NUs from the same direction of the sky as photon signals should be a robust way to identify sources, we pursue this approach for events that may be associated with NUs from the Galactic Center (GC).

A plausible source candidate for NU events with extreme energies is the SMBH at the dynamical center of our galaxy, Sagittarius (Sgr) A*.  Analysis of stellar orbits around Sgr A* demonstrate that the mass of this SMBH is about  $3 \times 10^6\, M_{\odot}$  \cite{Ghez:2008ms,Genzel:2010zy}.   Our GC is highly obscured. Sgr A* undergoes bursts of rapid variability in X-rays and gamma-rays; it is not visible in the optical and UV, and even in X-rays it is very dim.  The Fermi bubbles above and below the galactic plane were likely formed by jet activity in the distant past  \cite{Yang:2012fy}.  It has been suggested that the Fermi bubbles may be the origin of some IC events  \cite{Lunardini:2013gva}.

To the best of our knowledge, precise theoretical calculations of the expected NU flux from Sgr A*  have not been made.  High-energy protons from shock acceleration can interact with protons or photons to produce charged pions, kaons and neutrons that decay to NUs, along with neutral pions in the $pp$ channel that decay to photons.  Sgr A* is radiatively inefficient, so the photon density will be low in its environment. In SBGs, proton-proton interactions are expected to be dominant for PeV NU production  \cite{Laha:2013lka}. 

In Figure~\ref{skymap2}, we show that there are 9 IC NU events that are positionally consistent with the GC. We illustrate this using a dark (light) blue shaded region within $30^\circ$ ($45^\circ$) from the GC. In Table~\ref{table:GCevents}, we summarize the properties of these 9 IC events. In our subsequent statistical analyses, we will only consider the 7 IC events within a $30^\circ$ angular region of the GC (this excludes IC events \#12 and \#33).

%-------------------------------------------------------------------------------------------------------------------------------
% TABLE I
%-------------------------------------------------------------------------------------------------------------------------------
\begin{table}[htbp]
\hspace{-1.5cm}
\renewcommand{\arraystretch}{1}
\begin{center}
  \begin{tabular}{ | c | c | c | c | c | c | c | c |}
    \hline
      Event & Date & Energy & \hspace{5mm}RA\hspace{5mm} & \hspace{2mm}Dec.\hspace{2mm} & Pos.~Err. & Dis.~from GC & Other observations\\[-4mm]
 & (MJD) & (TeV) & (Deg) & (Deg)  &(Deg) & (Deg)  & \\
    \hline
    \hline
 2 & \hspace{5mm}55351.5\hspace{5mm} &  117 & 282.6 &-28&25.4& 14.6 & \emph{Swift} largest flare at 55359.5. \\[-4mm]
 & {\scriptsize Jun 4, 2010}&  & &&& & \\[-1mm]
% \cite{Degenaar:2012xh}.\\
    \hline
12 & 55739.4 & 104 &296.1 &-52.8&9.8& 32.5 & \emph{Swift} flare at 55739.5.\\[-4mm]
 & {\scriptsize Jun 27, 2011}&  & &&&  &\\[-1mm]

     \hline
14 & 55782.5 &1040 & 265.6&-27.9&13.2& 1.2 &  \emph{Swift} flare at 55790.4.  \\[-4mm]
 & {\scriptsize Aug 9, 2011} &  & &&& & \\[-1mm]

%\emph{Swift} observations within 3-4 days.
%INTEGRAL observations 2 days later. \\
    \hline
15 & 55783.2 & 57.5 & 287.3&-49.7&19.7& 26.3 &  \emph{Swift} flare at 55790.4.  \\[-4mm]
 & {\scriptsize Aug 10, 2011}&  & &&& & \\[-1mm]

%\emph{Swift} observations within 3-4 days.
%INTEGRAL observations 2 days later.\\
    \hline
22 & 55942.0 &  219.5 & 293.7 &-22.1&12.1& 25.9  & No X-ray observations.\\[-4mm]
 & {\scriptsize Jan 16, 2012}&  & &&&&  \\[-1mm]

    \hline
24 & 55950.8 & 30.5 &282.2 &-15.1&15.5& 20.4   & No X-ray observations.\\[-4mm]
 & {\scriptsize Jan 24, 2012}&  & &&& & \\[-1mm]

    \hline
25 & 55966.7 & 33.5 &286.0& -14.5&46.3& 23.5 & \emph{Chandra} largest flare at 55966.3.\\[-4mm]
 & {\scriptsize Feb 9, 2012} &  & &&& & \\[-1mm]

    \hline
33 &  56221.3 & 385 & 292.5&7.8&13.5& 44.8 &  \emph{Chandra} flare at 56222.7. \\[-4mm]
 & {\scriptsize Oct 21, 2012}&  & &&& & \\[-1mm]

%, but the positional coincidence shows a 2-3 sigma deviation.\\
    \hline
36 & 56308.2 & 28.9 & 257.7&-3.0&11.7& 27.2 & No X-ray observations. \\[-4mm]
 & {\scriptsize Jan 16, 2013} &  & &&& & \\

%Fermi flare 6 days earlier.
    \hline
  \end{tabular}
  \caption{Properties of the IC events consistent with a GC origin.  Our subsequent $p$-value analysis does not include IC events \#12 and \#33.  ``Pos. Err'' refers to the angular resolution of the IC measurement.}
  \label{table:GCevents}
\end{center}
\end{table}
%-------------------------------------------------------------------------------------------------------------------------------

%-------------------------------------------------------------------------------------------------------------------------------
\section{The Galactic Center: Sgr A* Flares}
\label{flares}
%-------------------------------------------------------------------------------------------------------------------------------

Targeted studies of Sgr A* have been made in gamma-rays by the \emph{Fermi} Large Area Telescope (LAT), in hard X-rays ($2-10$~keV) by the \emph{Chandra}~\cite{Baganoff:2001ju}, \emph{NuSTAR}~\cite{Barriere:2014bva}, and \emph{XMM-Newton} observatories and the \emph{Suzaku} satellite, in the NIR by the \emph{Hubble Space Telescope}, at multiple wavelengths (X-ray, optical, UV) by \emph{Swift}, and in the radio by the Very Large Array.  It is found that X-rays and NIR emission from Sgr A* have episodic flaring. Most of the time Sgr A* emits at low luminosity, but in its flares the brightness increases are a hundred-fold. A quiescent component dominates the emission at radio and submillimeter wavelengths.  It is the giant flares of Sgr A* that are of prime interest in seeing if there is an association with IC events, since the most energetic flares are the most likely to be associated with the HESE NUs, either as precursors or postcursors. We include X-ray flare information in the last column of Table~\ref{table:GCevents} and discuss the data sets below.

%-------------------------------------------------------------------------------------------------------------------------------
\subsection{\emph{Swift} X-ray Flares}
%-------------------------------------------------------------------------------------------------------------------------------

\begin{figure}[hbt]
\begin{center}
\includegraphics[width=0.53 \textwidth,trim=-.5 3.5  1.5  0,clip]{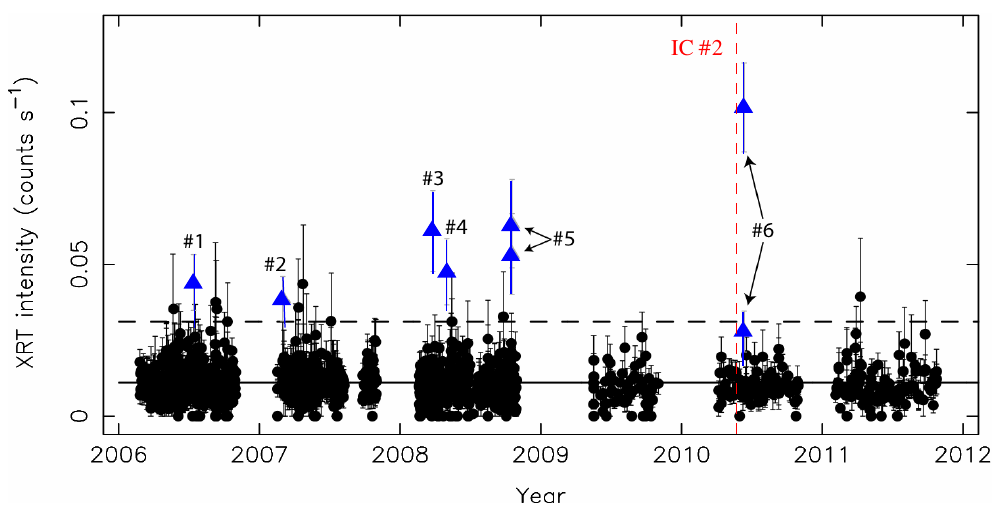}
\hspace{3mm}
\includegraphics[width=0.41 \textwidth,trim=-0 -3  0  0,clip]{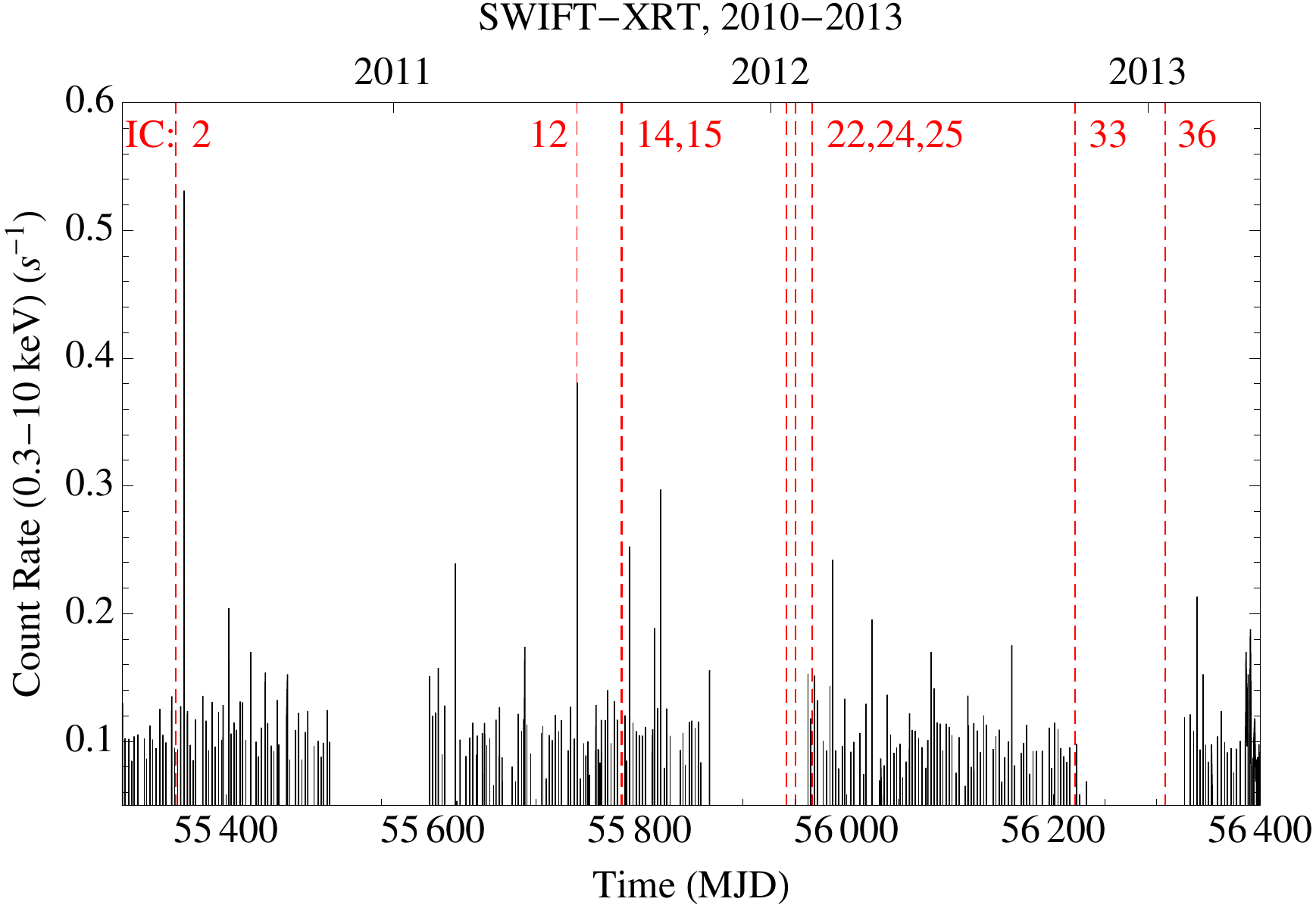}
\caption{\label{swift} Left: \emph{Swift} XRT observations of Sgr A* over six years \cite{Degenaar:2012xh}. The events numbered  \#1,  \#2, ...  \#6 in blue are \emph{Swift} flares.  The red dashed line shows the timing of IC \#2. Right: \emph{Swift} XRT observations during the IC observation period, with the times of potential GC NU events marked with red dashed lines. The X-ray data were generated using \emph{Swift}'s online tools, see \cite{Evans:2007na}.}
\end{center}
\end{figure}

In Figure~\ref{swift}, we show the \emph{Swift} X-Ray Telescope (XRT) observations of Sgr A* versus time. \emph{Swift} detected 6 hard X-ray flares from Sgr A* during six years of intermittent observations, constraining the occurrence rate of bright ($L_X>10^{35}$~erg~s$^{-1}$) X-ray flares to be $\sim0.1-0.2$ per day \cite{Degenaar:2012xh}.  The flares occurred close to the Sgr A* SMBH event horizon, as inferred from both the total time duration and the short timescale variability.  What powers the flares is unknown. Interestingly, the largest flare, \#6, occurred near to the time of IC event \#2 (i.e., the flare happened 8 days after the IC event). IC event \#12 also has a time match to a flare.

%-------------------------------------------------------------------------------------------------------------------------------
% FIGURE 3
%-------------------------------------------------------------------------------------------------------------------------------
%-------------------------------------------------------------------------------------------------------------------------------

%-------------------------------------------------------------------------------------------------------------------------------
\subsection{\emph{Chandra} and \emph{NuSTAR} X-ray Flares}
%-------------------------------------------------------------------------------------------------------------------------------

\begin{figure}[hbt]
\begin{center}
\includegraphics[width=0.43 \textwidth,trim=0 -10 -15 -10,clip]{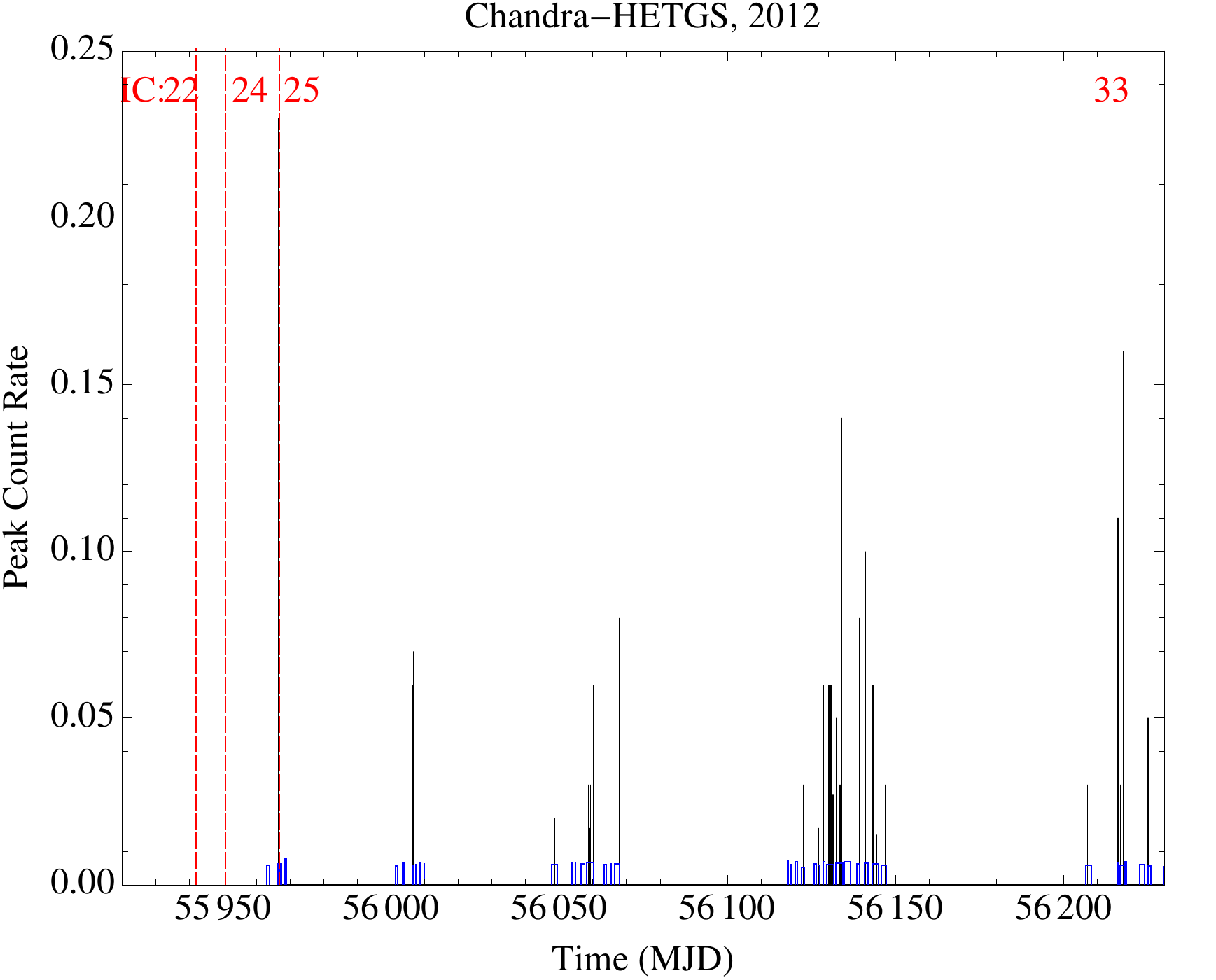} \hspace{1mm}
\includegraphics[width=0.55 \textwidth,trim=30 150 45 320,clip]{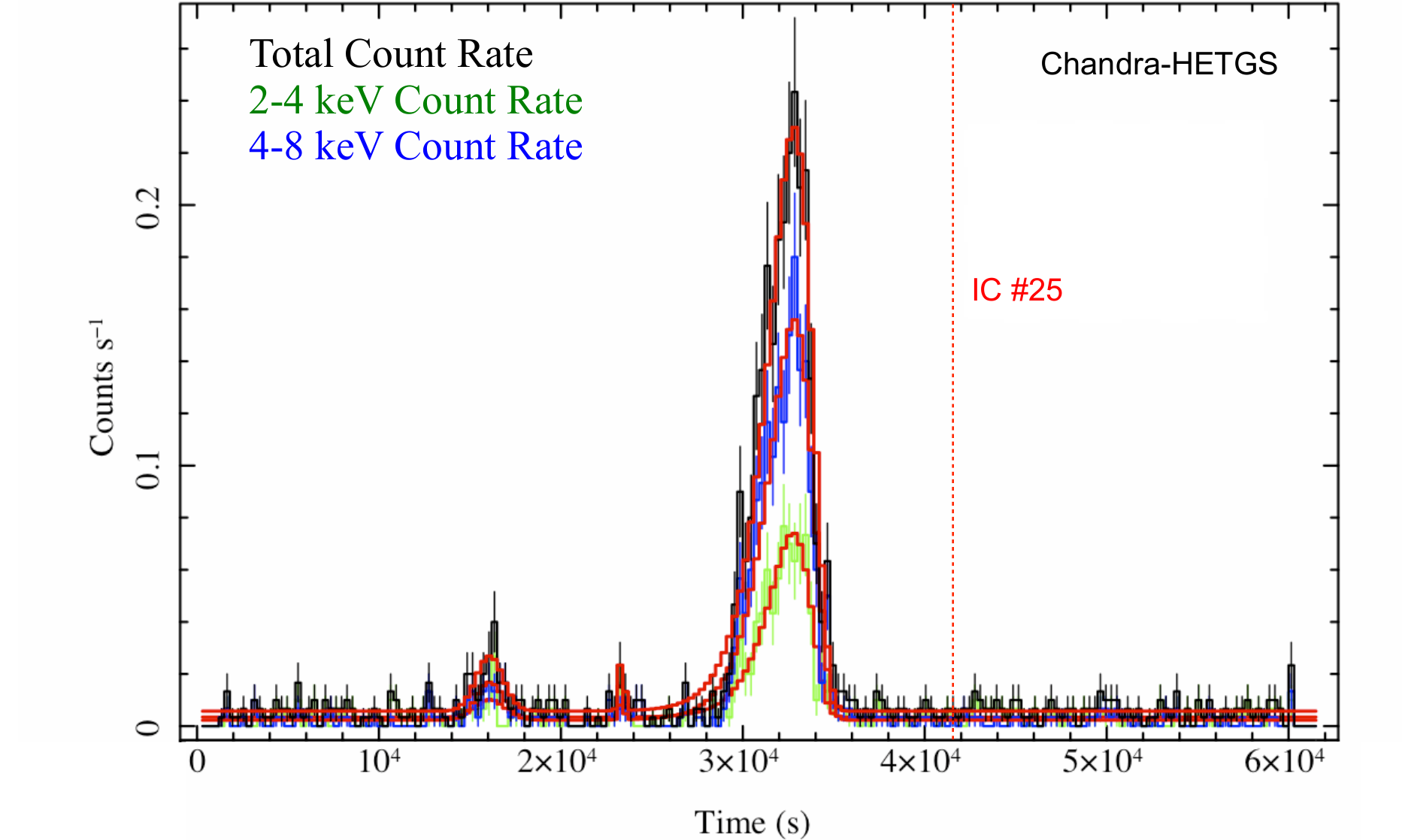}
\caption{\label{chandra} Left: \emph{Chandra} HETGS observations of Sgr A* in 2012  \cite{Neilsen:2013sha}. The blue regions along the bottom are the quiescent levels and show when \emph{Chandra} observations were being made. Right: A close-up on the largest flare, observed February 9th  \cite{Nowak:2012ry}. The red dashed line shows the timing of IC  \#25.  IC \#25 occurred 2:38:29 hours after the \emph{Chandra} giant flare.}
\end{center}
\end{figure}

In February 2012, \emph{Chandra} began a dedicated 3~Ms observational program of Sgr A* \cite{Neilsen:2013jja} using the High Energy Transmission Grating Spectrometer (HETGS).  The goals of this \emph{Chandra} X-ray Visionary Project (XVP) were to study the physics of X-ray flares with the highest spatial and spectral X-ray resolution available and to investigate their relationship to the quiescent X-ray emission. During the XVP, 39 flares were identified. A typical flare was observed to be about 10 times brighter than the background emission.  We show these observations in the left panel of Figure~\ref{chandra}.

The brightest Sgr A* X-ray flare ever observed occurred on February 9, 2012 (observation ID 14392). Its peak $2-10$~keV luminosity was $L_X \sim 5\times10^{35}$~erg~s$^{-1}$.  The duration of the flare was 5.9~ks, and its peak flux occurred at 15:10:21 UTC  \cite{Nowak:2012ry}. Intriguingly, IC $ \#$25 happened on the same date.  It was a shower event with an energy of 33.5~TeV and a direction consistent with a GC origin within its $1\sigma$ uncertainty (see Figure \ref{skymap}).  The recorded time of this IC event was 55966.7422457 MJD = 17:48:50 UTC.  Thus, IC $ \#$25 occurred about 2:38:29 hours after the X-ray flare (see right panel of Figure \ref{chandra}).  Because of the complex dynamics of the processes at the GC with X-rays of electromagnetic origin and NUs of hadronic origin, the somewhat inexact time coincidence of the X-ray activity and NU signal may not be surprising. The association of large X-ray flares with IC events is serendipitous, and they may not always occur in conjunction. 

%-------------------------------------------------------------------------------------------------------------------------------
% FIGURE 4
%-------------------------------------------------------------------------------------------------------------------------------
%-------------------------------------------------------------------------------------------------------------------------------

The \emph{NuSTAR} high-energy X-ray observatory observed the GC three times in July, August, and October 2012~\cite{Barriere:2014bva} as part of a coordinated campaign with \emph{Chandra} and the Keck Telescope.  Four flares were observed by \emph{NuSTAR}, two of medium amplitude and two weaker ones.  One of these flares was observed on October 17, four days preceding IC event \#33. This flare was simultaneously observed by \emph{Chandra}.

%----------------------------------------------------------------------------------------------------------------------------
\subsection{\emph{Fermi} Activity}
%-------------------------------------------------------------------------------------------------------------------------------

\begin{figure}[htb]
\begin{center}
\includegraphics[width=0.69 \textwidth]{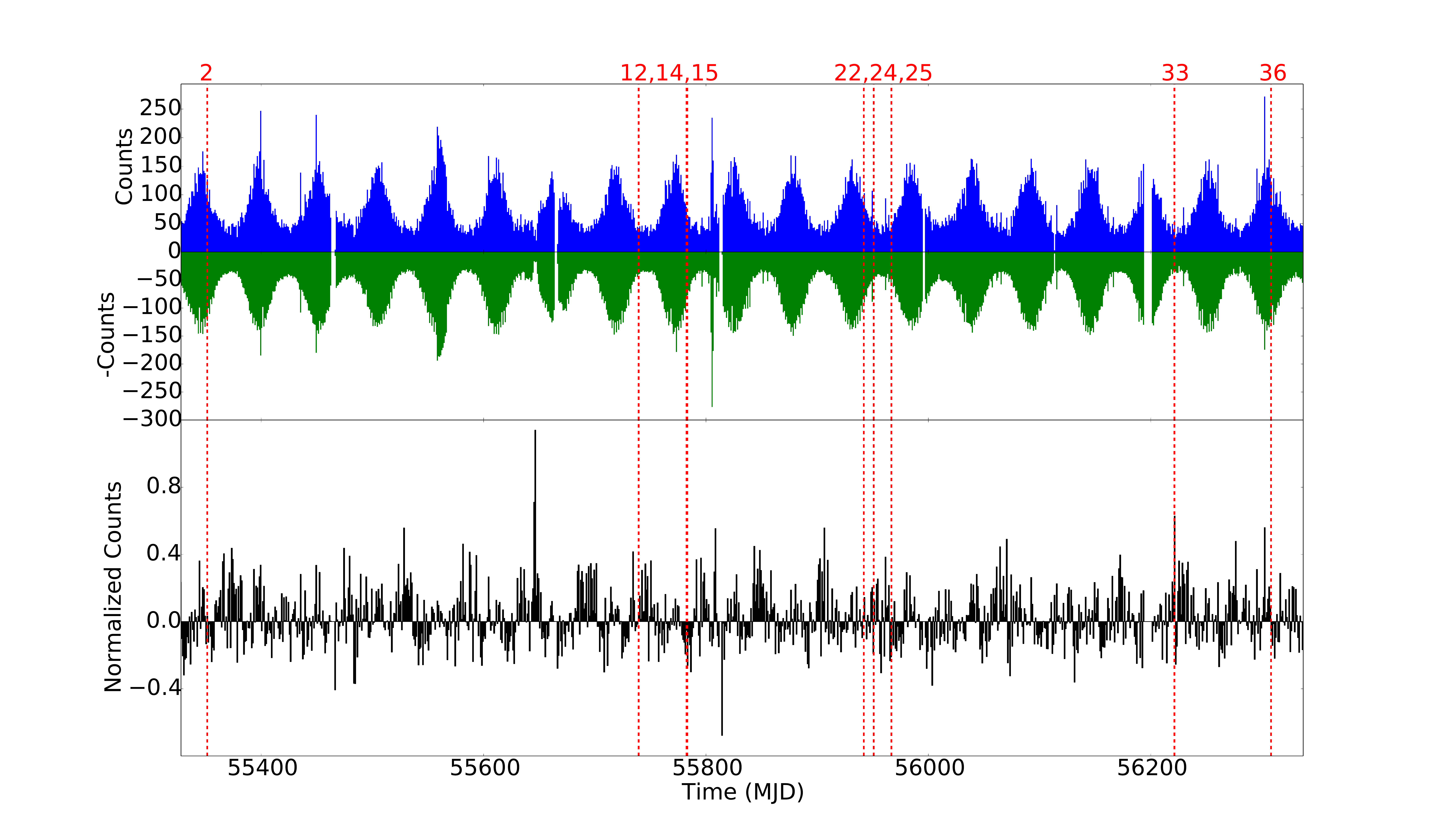}
\caption{\label{fermi} \emph{Fermi} observations of Sgr A* within a $1^\circ$ cone. The red dashed lines show the timing of IC NU events. To identify activity above quiescent levels, we simulated the expected data (assuming Sgr A* is a steady source) and subtracted the simulated data (green) from the observed data (blue). We show the normalized difference in the bottom panel.}
\end{center}
\end{figure}

The \emph{Fermi} telescope has made gamma-ray observations of the GC. In order to identify flare activities in their data, we used the ObsSim program from the \emph{Fermi} SciTools to obtain the satellite position data and to generate a background distribution. Basically, we specified a point source at the GC with some arbitrary luminosity and power-law spectrum. We then used the simulation program to read the \emph{Fermi} satellite data~\cite{FermiLATdata} and to generate photon events from this source when \emph{Fermi} LAT is taking data. The simulated events take into account the orientation of the satellite and assume a $1^\circ$ cone of observation. We also normalized the simulated events to the total observed number of events. In Figure \ref{fermi}, we show the difference between the data and a simulated background that assumes a steady source at Sgr A*. The times of the IC events in the GC region are indicated by the red dashed lines.  No appreciable flaring is observed.  This is not so surprising, given that a coincidence between the IC NU events and GRBs has not been found \cite{Aartsen:2013jdh, Aartsen:2014gkd, Casey:2013, Yacobi:2014vja}.

%-------------------------------------------------------------------------------------------------------------------------------
% FIGURE 5
%-------------------------------------------------------------------------------------------------------------------------------
%-------------------------------------------------------------------------------------------------------------------------------

%-------------------------------------------------------------------------------------------------------------------------------
\section{Time Clustering of Neutrino Events from the Galactic Center}
\label{clustering}
%-------------------------------------------------------------------------------------------------------------------------------

\begin{figure}[h!]
\includegraphics[width=0.45\textwidth,trim=0 -10 -25 -10,clip]{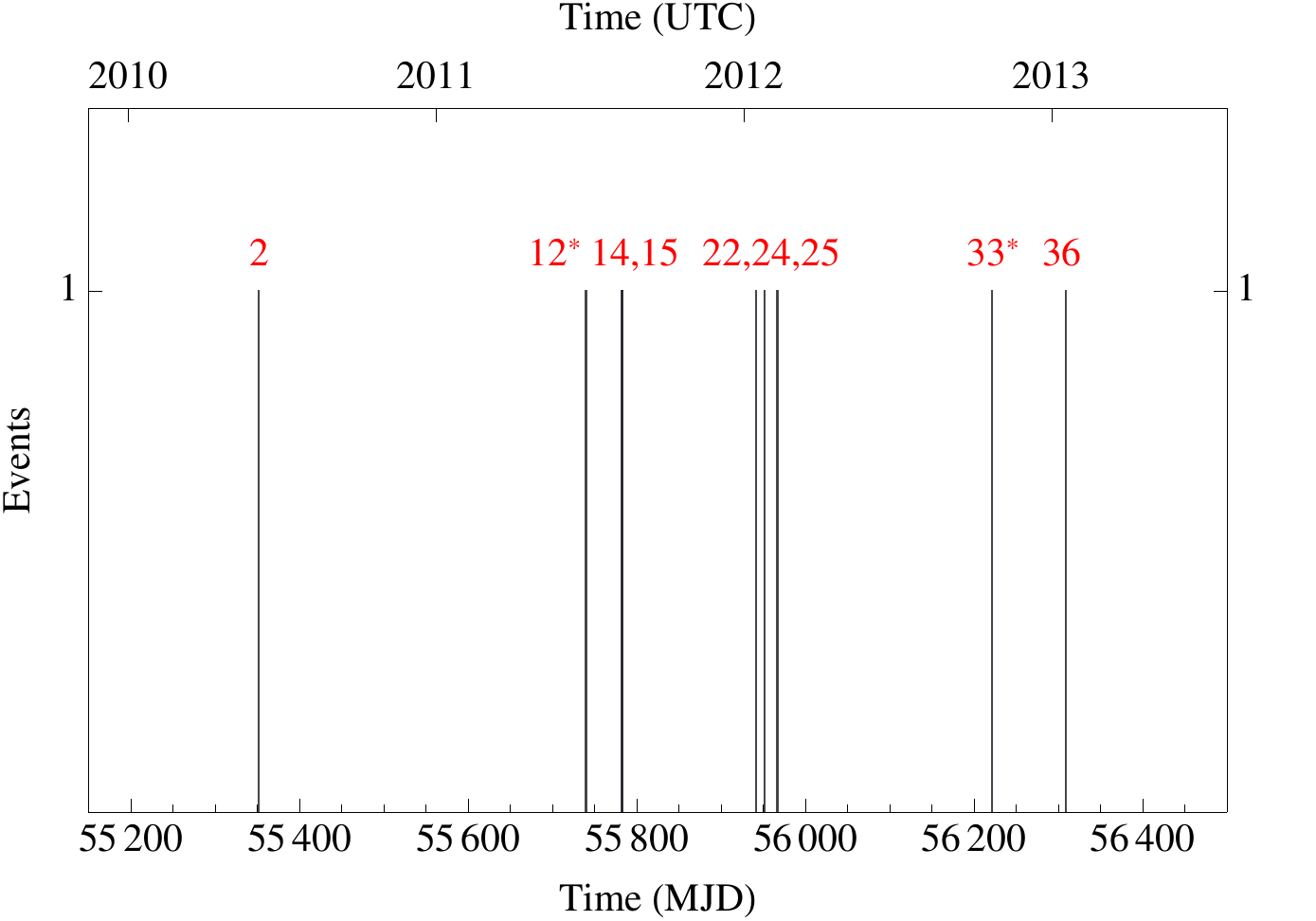} \hspace{2mm}
\includegraphics[width=0.45\textwidth,trim=0 -10 -25 -10,clip]{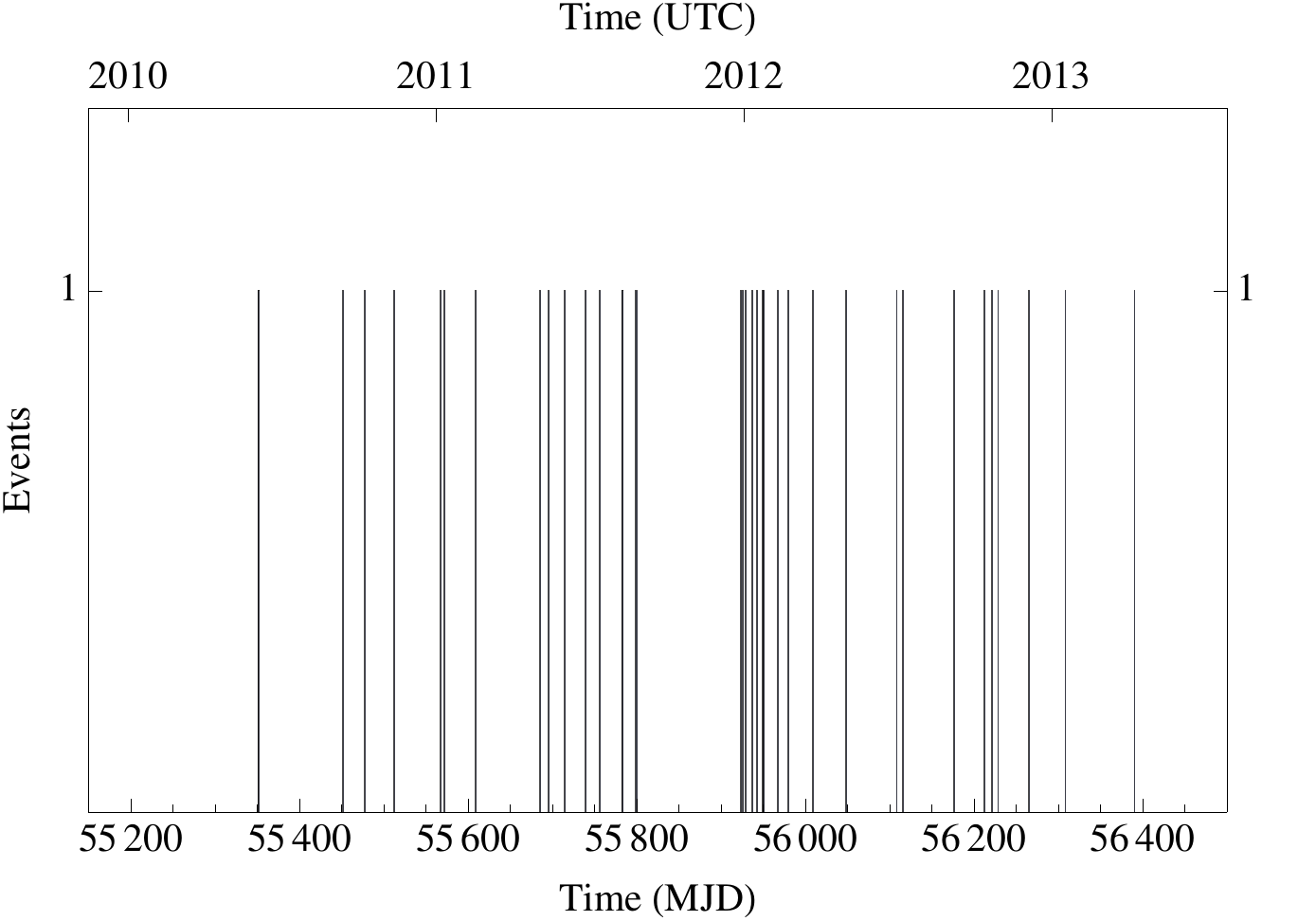}
\caption{\label{timing} Time sequence of IC events that are positionally consistent within $45^\circ$ of the GC. Events \#12 and \#33 are more than $30^\circ$ from the GC. For comparison, the time sequence of all IC events is also shown.}
\end{figure}

\begin{figure}[!htb]
\centering
\includegraphics[scale=.5]{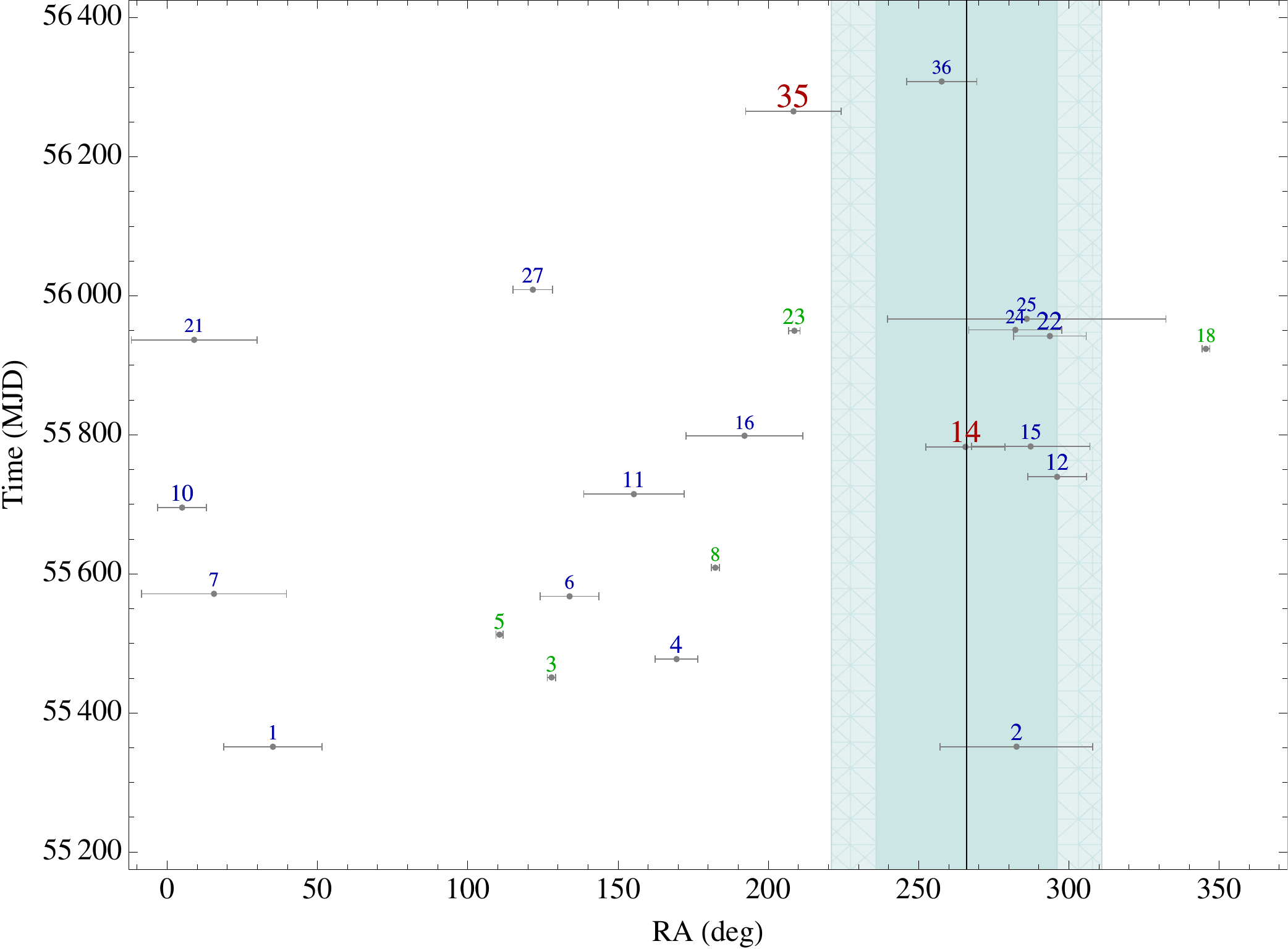}
\caption{The time and RA of all IC events within $30^\circ$ of declination of the GC.  The inner [outer] blue band has $\mbox{RA} \in (236^\circ, 296^\circ) = (\mbox{RA}_{\rm GC}-30^\circ,\mbox{RA}_{\rm GC}+30^\circ)$ [$\mbox{RA} \in (221^\circ,311^\circ) = (\mbox{RA}_{\rm GC}-45^\circ,\mbox{RA}_{\rm GC}+45^\circ)$]. Events positionally consistent within $30^\circ$ of the GC, which fall within the inner blue band, show more time clustering than events away from the GC. }
\label{fig:timecorr}
\end{figure}

Time clustering of the IC NU events in the GC can indicate a transient nature of the underlying physics. In the left panel of Figure~\ref{timing}, we plot in time sequence the 9 IC events that are positionally consistent within $45^\circ$ from the GC.  A visual comparison of this plot with all the IC events (right panel of Figure~\ref{timing}) suggests the occurrence of NU bursts.  We hereafter focus on the 7 IC events  within $30^\circ$ from the GC.  We evaluate the probability ($p$-value) that the 7 IC events are not clustered, i.e., that they are randomly distributed in time. In Figure \ref{fig:timecorr}, we show both time and space clustering of IC events. One can easily see that IC events \#22, \#24, and \#25 are clustered both in timing and in location, as are \#14 and \#15.

%-------------------------------------------------------------------------------------------------------------------------------
% FIGURE 6
%-------------------------------------------------------------------------------------------------------------------------------
%-------------------------------------------------------------------------------------------------------------------------------

%-------------------------------------------------------------------------------------------------------------------------------
% FIGURE 7
%-------------------------------------------------------------------------------------------------------------------------------
%-------------------------------------------------------------------------------------------------------------------------------

%-------------------------------------------------------------------------------------------------------------------------------
\subsection{Self-Clustering Analysis}
%-------------------------------------------------------------------------------------------------------------------------------

IC has made studies of the timing correlations of various subsets of the IC events~\cite{Braun:2008bg, Braun:2009wp, Aartsen:2014gkd, Aartsen:2014cva}.  We adopt the IC collaboration methodology for our analysis of the 7 IC events that are less than $30^{\circ}$ from the GC: IC events \#2,\#14,\#15,\#22,\#24,\#25, and \#36. For every pair of events, with times $t_{\rm left}$ and $t_{\rm right}$, we define a signal function for the IC event \#$i$ as
\begin{equation}
S_i=\frac{H(t_{\rm right}-t_i)H(t_i-t_{\rm left})}{t_{\rm right}-t_{\rm left}} \,,
\end{equation}
and a background function as
\begin{equation}
B=\frac{1}{T} \,,
\end{equation}
where $H$ is the Heaviside function and $T=998$~days  is the total observation time.
We then define a likelihood function
\begin{equation}
\mathcal{L}=\prod_{i\in {\rm events}}\left( \frac{n_s}{N_{\rm events}}S_i+\frac{N_{\rm events}-n_s}{N_{\rm events}}B\right) \, .
\end{equation}
Here, $n_s$ is the number of signal events in a cluster, and $N_{\rm events}=7$ is the total number of events. In order to compute the test statistics (TS), we marginalize $n_s$ and choose the pair of events that gives the best TS. We generate events randomly over the total observation time and marginalize over $n_s$ to compute the $p$-value. We find a $p$-value of 1.6\% for the pair of IC events \#14 and \#15 ($n_s=2$; $\Delta t = 0.67$~days).

%Likelihood clustering P-value:
%Using Icecube approach
%Events:  2,14,15,22,24,25,33
%P-value 1.6\%
%$Delta_t = 0.67$   size of the cluster.
%Events = 14,15  Events involved in the cluster.
%$n_s$ best = 2      Number of signal events.

%-------------------------------------------------------------------------------------------------------------------------------
\subsection{Friends-of-Friends Clustering Analysis}
%-------------------------------------------------------------------------------------------------------------------------------

The above TS analysis found that the clustering of IC events \#14 and \#15 happens with a probability of $p = 1.6\%$ compared to events that are randomly distributed in time. This result may indicate that the two events come from the same transient phenomenon, but it does not check the clustering of all the GC IC events. To test the latter,  we use the Friends-of-Friends algorithm \cite{Zeldovich:1982zz}.

The algorithm consists of grouping events together if they are friends or connected by friends:  two events are friends if they are closer than some threshold
distance $\delta t_{\rm friends}$.  We define the TS to be the minimum $\delta t_{\rm friends}$ that we need to form a given number of clusters.
For randomly generated events, we obtain a minimum $p$-value of 4.2\% with the following 4 clusters of IC events:  (\#2), (\#14, \#15), (\#22, \#24, \#25), (\#36).

%-------------------------------------------------------------------------------------------------------------------------------
\subsection{Likelihood Analysis for \emph{Chandra} Flare Coincidence with IC Neutrino Events}
%-------------------------------------------------------------------------------------------------------------------------------

To evaluate the probability of a random coincidence of \emph{Chandra} X-ray flares with the 7 IC events within $30^\circ$ from the GC, we perform a likelihood test with the 37 X-ray flares observed at the GC by \emph{Chandra}.  We define the signal function as a top-hat distribution around the flare with a time window $\Delta t$ that is weighted with the counts in the flare.  The events are distributed over a total time of  $T=3$~Ms. For the background function, we take a flat distribution over all observation time periods that is normalized to the total number of flare counts. As in the time clustering analysis above, we marginalize $n_s$ to calculate the TS values. We do the same with randomly generated events during the total observation time to compute the $p$-value. Based on the duration of the observed Sgr A* flares, we choose a time window of $\Delta t=12$ hours, for which we obtain $p=0.9\%$ with $n_s = 1$.

%-------------------------------------------------------------------------------------------------------------------------------
\section{Association of TeV-PeV GC Gamma Rays with PeV GC Neutrinos}
\label{gammaray}
%-------------------------------------------------------------------------------------------------------------------------------

As discussed in the Introduction, the pion production in $pp$ interactions gives both NUs (from charged pion decays) and gamma rays (from neutral pion decays). Thus, the existence of the new physics NU signal implies the existence of high-energy gamma rays. Our argument that NUs are produced at the GC then necessitates a high-energy gamma ray signal at about the same rate, and this prediction can validate or rule out our hypothesis. In the following, we elaborate on and quantify the gamma ray prediction.
 
The acceleration of protons and nuclei by the Fermi mechanism gives CRs with a $E^{-2}$ spectrum, $dN/dE \propto E^{-2}$.  The hadronic interactions of these energetic CR with the diffuse gas surrounding the cosmic accelerator produce mesons (pions, kaons, charm), whose energy spectra are slightly softer, $E^{-2.3}$, than the primary CRs.  See, e.g. \cite{Dar:2014, Halzen:2013xxa}. The inelastic $pp$ collisions populate a democratic pion multiplicity with about equal numbers of $\pi^+$, $\pi^-$, and $\pi^0$.  The two-body decays, $\pi^+ \to \mu^+  \nu_\mu$, $\pi^- \to \mu^- \bar{\nu}_\mu$, $\pi^0 \to 2 \gamma$, have primary NUs and photons with similar distributions in neutrino and photon energies. In the propagation of the NUs over long baselines the initial flavor converts to a 1:1:1 composition of $\nu_e, \nu_\mu, \nu_\tau$. Consequently, the ratio of the photon to $\nu$
distributions (at $E_{\gamma} = E_{\nu}$) is 
\beq
\frac{dN}{dE_\gamma} ={\cal O}(1)\times \frac{dN}{dE_\nu}  \,,
\eeq
which provides a correlated prediction of high-energy gamma ray flux from the NU flux~\cite{Ahlers:2013xia, Supanitsky:2013ooa}.

%-------------------------------------------------------------------------------------------------------------------------------
% FIGURE 8
%-------------------------------------------------------------------------------------------------------------------------------
\begin{figure}[!htb]
\centering
\includegraphics[width=0.6\textwidth]{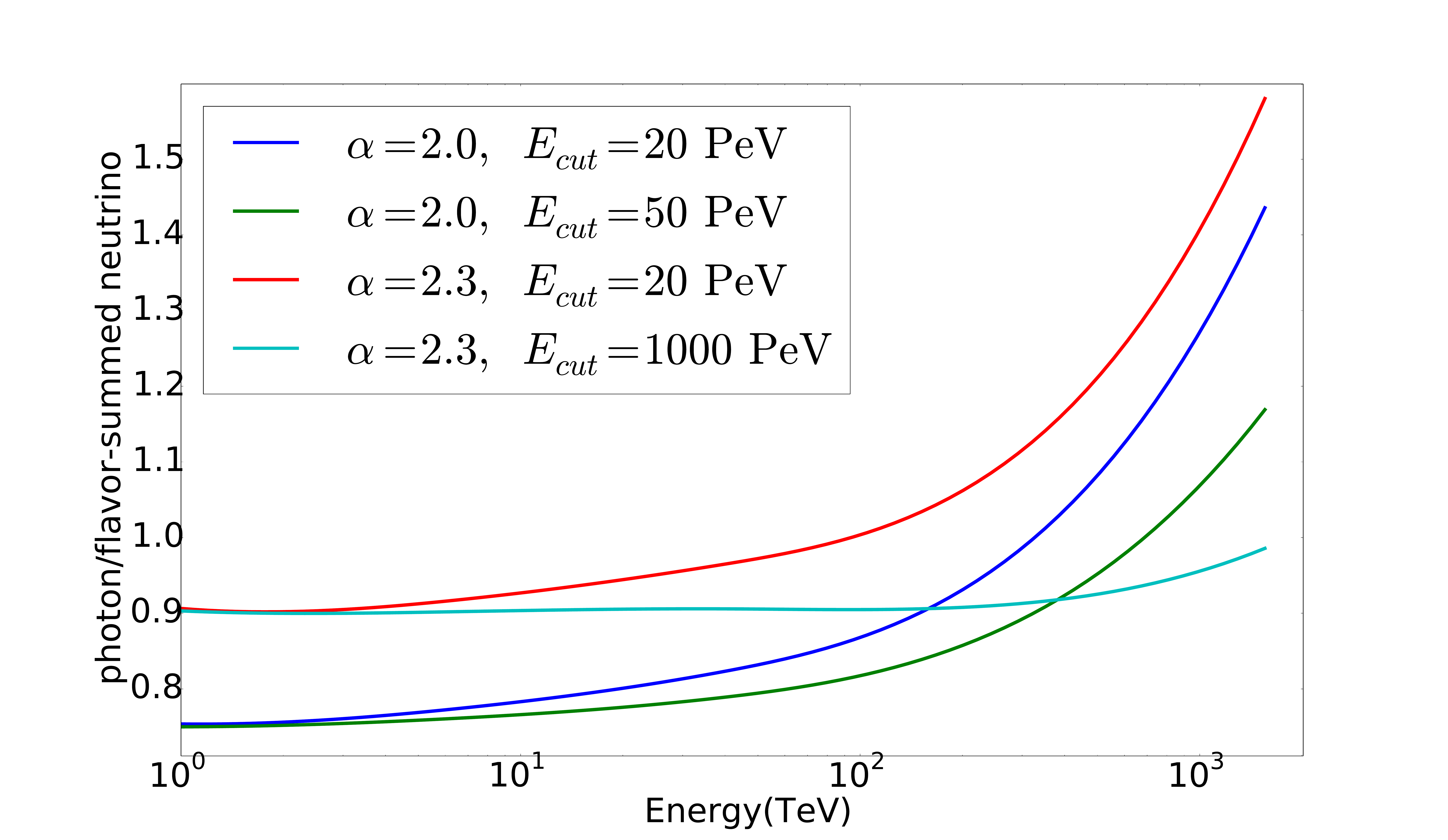}
\caption{The ratio of the gamma ray flux to the flavor-summed neutrino flux as a function of energy. See Eq.~(\ref{eq:formula-proton}) for the definition of $\alpha$ and $E_{\rm cut}$.}
\label{fig:photon-neutrino-ratio}
\end{figure}
%-------------------------------------------------------------------------------------------------------------------------------

We use Pythia8~\cite{Sjostrand:2007gs} to simulate the productions of neutrino and photon from the inelastic scattering of a cosmic ray proton with a proton at rest. We use the following spectrum formula for the primary proton flux
\beq
 \frac{dN(E_p)}{dE_p} = A_p \, E_p^{- \alpha} e^{-E_p/E_{\rm cut} } \,,
 \label{eq:formula-proton}
\eeq
with a power law below the cutoff $E_{\rm cut}$. Here, $A_p$ is an energy-independent normalization factor. To match the observed neutrino spectrum at IC, we choose $\alpha = 2.0$ and $\alpha=2.3$ as two representative powers. In Figure~\ref{fig:photon-neutrino-ratio}, we show the ratio of the gamma ray flux to the neutrino flux as a function of energy. For a very large cutoff $E_{\rm cut}=1000$~TeV, the photon over proton flux ratio is a constant for energies below 100 TeV. However, for a smaller value of cutoff, the the photon/neutrino ratio increases as the energy approaches the cutoff. This can be understood from kinematics of charged pion decaying into NUs and neutral pion decaying into two photons.

\begin{figure}[tbh!]
\centering
\includegraphics[width=0.6\textwidth]{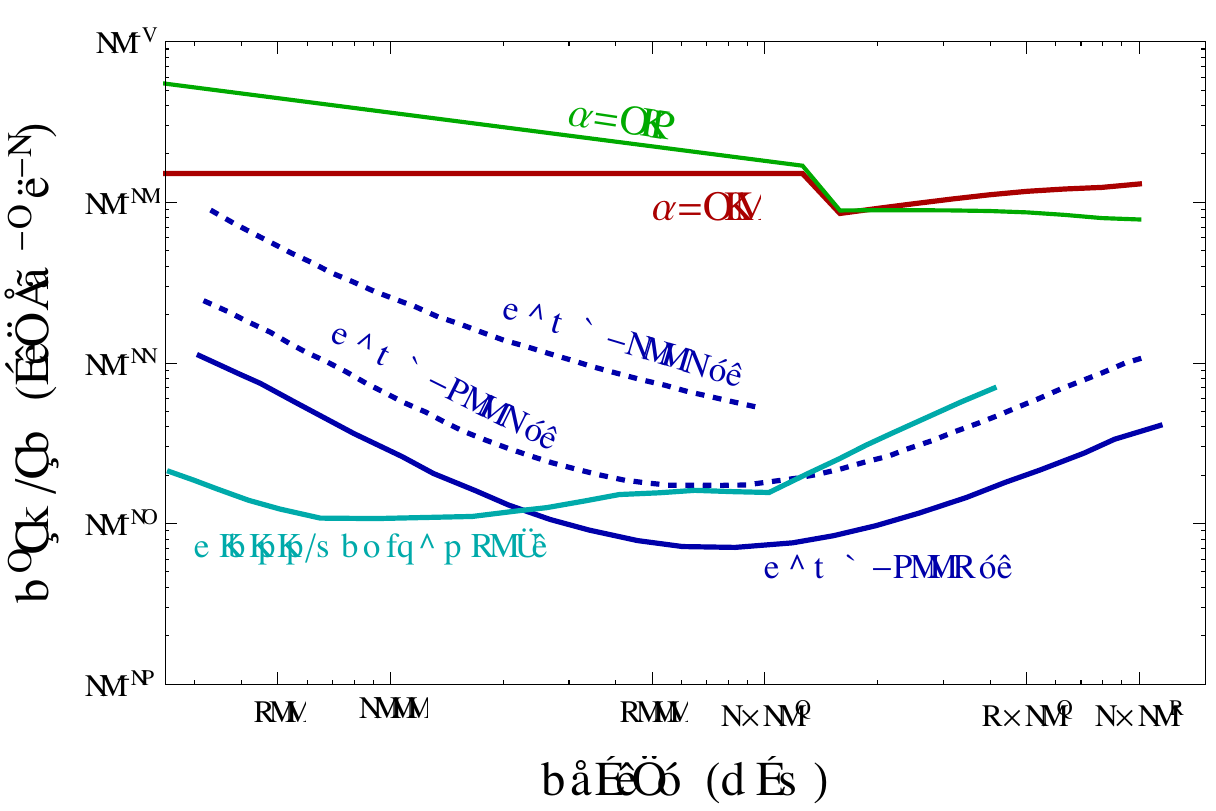}
\caption{The correlated gamma ray flux from the IC NU flux based on $pp$ collisions. The predicted values of the gamma flux are shown by the green and red curves based on primary proton spectra with a power-law, $E^{-\alpha}$, with $\alpha=2.3$ and $\alpha=2.0$, respectively. The breaks in the predicted gamma ray flux are due to the attenuation effects. Here, the unit $1~\mbox{erg} = 624.15$~GeV. Future experimental sensitivities are taken from Ref.~\cite{Abeysekara:2013tza}.}
\label{fig:photon-flux}
\end{figure}

The high energy gamma rays with $E \gtrsim 10$~TeV in Sgr A* have attenuation effects when gamma rays interact with Galactic interstellar radiation field and CMB photons by pair production. The CMB photon effects become more important for the gamma ray energy above around 200 TeV. One can use the optical depth $\tau_{\gamma \gamma}$ to quantify the attenuation effect such that 
\beq
\frac{dN(E_\gamma)}{d E_\gamma}\Big|_{\oplus} = \frac{dN(E_\gamma)}{d E_\gamma}\Big|_{\rm GC} \times e^{-\tau_{\gamma \gamma} (E_\gamma)} \,.
\eeq
We use the following numerical function to fit the calculated optical depth in Ref.~\cite{Moskalenko:2005ng}
\beq
\tau_{\gamma \gamma}(E_\gamma) \,=\,\sum_{i=0}^{10} a_i  \left( \log_{10}{E_\gamma} \right)^i \,,
\eeq
for $14~\mbox{TeV} < E_\gamma \leq 1000$~TeV and $\tau_{\gamma \gamma}(E_\gamma)=0$ for $E\leq 14~\mbox{TeV}$. The fitted parameter values are
\beqa
(a_0, a_1, a_2, \cdots, a_{10})&=& (-553.195, 2368.8, -4399.76, 4664.28, -3125.54, 1384.65,  \nonumber \\  
&& \hspace{1cm} -411.397,  81.1285, -10.1884, 0.737702, -0.0234442) \,.
\eeqa
After folding the attenuation effects, we show the predicted gamma ray flux in Figure~\ref{fig:photon-flux} for the range of energy from a few hundred GeV to 200 TeV. The optical depth used in the above formula is just an averaged one. If the local environment of the source to generate IC NUs has a large value of optical depth, an even smaller gamma ray flux will be expected.

%-------------------------------------------------------------------------------------------------------------------------------
% FIGURE 9
%-------------------------------------------------------------------------------------------------------------------------------
%-------------------------------------------------------------------------------------------------------------------------------

New-generation ground-based gamma ray observatories are coming into operation that will probe high energy gamma rays \cite{Holder:2014eja}.  The High Altitude Water Cherenkov (HAWC) observatory in Mexico is sensitive to gamma rays and cosmic rays of 100 GeV to a few hundred TeV \cite{Abeysekara:2013tza, Abeysekara:2014ffg}.  HAWC has an all-sky field of view and nearly continuous operation. We show the projected sensitivities in Figure~\ref{fig:photon-flux}. The Cherenkov Telescope Array (CTA) \cite{Doro:2012xx,Pierre:2014tra}, with tens of telescopes in several sites, will provide energy coverage from tens of GeV to several tens of TeV.  Its sensitivity will be about a factor of 10 better than the current H.E.S.S., MAGIC, MILAGRO and VERITAS gamma ray detectors.  CTA will have a field of view of up to 10 degrees and also have sensitivity to confirm the hadro-production of NUs around Sgr A*~\cite{Linden:2012bp}.

%-------------------------------------------------------------------------------------------------------------------------------
\section{Conclusions}
\label{conclusions}
%-------------------------------------------------------------------------------------------------------------------------------

We proposed that the timings of IC NU events from Sgr A* are sometimes correlated with the observed photon flaring in X-rays at the GC.  In particular, we consider the timing and approximate positional coincidences of IC \#25 and \emph{Chandra} \#14392 as an indicator that Sgr A* is the source of IC \#25. A testable consequence of this interpretation of the data is that major photon flares of other AGNs (or of the cores of SBGs) occur simultaneously with extragalactic IC events. This conclusion implies that hadronic processes produce NUs from pion production and their subsequent decays, along with the Inverse-Compton mediated X-ray flaring. We also investigated the idea that NU bursts from Sgr A* occur.  We found support for this hypothesis in the low probability of random emissions in time to explain the IC observations.  

X-ray observations of Sgr A* will continue to provide valuable information about the frequency and brightness of the flares.  IC continues to map the NU sky at the highest NU energies.  Further coincidences in the timing of X-ray flares with IC events that point to the GC would bolster the AGN point-source connection.  Our expectation is that HESE NUs would occur in association with giant X-ray and NIR flares \cite{Razzaque:2013uoa}.  In addition, other variable sources in the GC region near Sgr A*, such as the transient magnetar SGR J1745-29 (e.g.~\cite{Rea:2013pqa,Degenaar:2014qoa}), could be another source of NUs.

Unusual outbursts could serendipitously arise from disruptions of asteroids, comets, planets or stars that approach the SMBH.  A gas cloud of Earth-mass, called G2, is approaching Sgr A*  \cite{DoddsEden:2010wx,Moscibrodzka:2012sg}. The trajectory of G2 is predicted to reach the pericenter of the orbit in 2014.  It will be especially interesting if any high-energy NU events from Sgr A* occur that can be associated with its passage.

%-------------------------------------------------------------------------------------------------------------------------------
{\bf Acknowledgments:} \ We thank Philip Armitage, Francis Halzen and Albrecht Karle for informative discussions. The work is supported by the U. S. Department of Energy under the contract DE-FG-02-95ER40896, the U. S. National Science Foundation, NASA and the David and Lucille Packard Foundation. YB thanks the Center for Future High Energy Physics and the Aspen Center for Physics, under NSF Grant No. PHY-1066293, where this work is finished.
%-------------------------------------------------------------------------------------------------------------------------------

\providecommand{\href}[2]{#2}\begingroup\raggedright\endgroup

%-------------------------------------------------------------------------------------------------------------------------------

\end{document}